\documentclass[prb,aps,amsmath,showpacs,twocolumn]{revtex4}

\usepackage{graphicx}% Include figure files

\usepackage{dcolumn}% Align table columns on decimal point

\usepackage{bm}% bold math

\begin{document}

\title{Electronic Structure of Samarium Monopnictides and Monochalcogenides}
\author{A. Svane$^1$, V. Kanchana$^2$, G. Vaitheeswaran$^2$, G.  Santi$^1$, \\
W. M. Temmerman$^3$, Z. Szotek$^3$, P. Strange$^4$ and L. Petit$^{1,5}$}
\affiliation{$^1$Department of Physics and Astronomy, \\
University of Aarhus, DK-8000 Aarhus C, Denmark}
\affiliation{$^{2}$Max-Planck-Institut f\"ur Festk\"orperforschung, 
D-70569 Stuttgart, Germany}
\affiliation{$^{3}$Daresbury Laboratory, Daresbury, Warrington WA4 4AD, United Kingdom}
\affiliation{$^{4}$Department of Physics, Keele University, ST5 5DY, United Kingdom }
\affiliation{$^{5}$Computer Science and Mathematics Division, and Center for
Computational \\
Sciences, Oak Ridge National Laboratory, Oak Ridge, TN 37831, USA}
\date{\today}

\begin{abstract}
The electronic structures of SmX (X=N, P, As, Sb, Bi, O, S, Se, Te, Po) compounds are
calculated using the self-interaction corrected local-spin density approximation.
The Sm ion is described with either five or six localized $f$-electrons while the
remaining electrons form bands, and the
total energies of these scenarios are compared. 
With five localized $f$-electrons a narrow $f$-band is formed in the vicinity of
the Fermi level leading to an effective intermediate valence. This scenario is
the ground state of all the pnictides as well as SmO. With six localized $f$-electrons,
the chalcogenides are semiconductors, which is the ground state of SmS, SmSe and SmTe.
Under compression the Sm chalcogenides undergo 
first order transitions with destabilization of the $f$ states into the intermediate
valence state, the bonding properties of which are 
 well reproduced by the present theory.
\end{abstract}

\pacs{71.20.Eh, 71.28.+d, 71.15.Mb, 71.15.Nc}
\maketitle

\section{Introduction}

The valency of rare earth compounds continues 
to be a vivid research 
area.\cite{wachter,LeBihan,Schumann,Lehner,nature,Antonov,Chainani,Raymond,Barla} 
Of particular interest are
systems where valency is influenced by controllable 
external parameters like 
pressure, temperature or alloying. Several
 systems with varying or fluctuating valencies 
are known.\cite{wachter,Boppart} One of the most studied systems 
is SmS,\cite{wachter,Jayaraman,Varma,Varma1}
which at low temperature and zero 
pressure crystallizes in the NaCl structure 
with a semiconducting behavior, in
accordance with the presence of divalent Sm ions and a full S $p$-band. At 
a moderate pressure of $\sim 0.65$ GPa  SmS reverts 
to a metallic
phase with a significant volume collapse of 13.5\%,\cite{Benedict1} however retaining the
NaCl structure. At very low temperature, the high pressure phase in fact 
reveals a small pseudogap, of the order of $\sim 7$ meV.\cite{wachter}
%This phenomenon can be attributed to a valence
%transition of Sm ion from divalent to trivalent state
%within the fcc phase and the transition is of first order\cite{Jayaraman}.
Photoemission experiments show distinctly different spectra
for the two phases, which usually are 
interpreted on the basis of divalent $f^6$ ions 
in the ground state and mixed valent
$f^5-f^6$ ions in the high pressure metallic phase.\cite{Campagna,Oh} 
Recent experiments also find indications of
small amounts of $f^5$ character in the ground state.\cite{Chainani}
The $C_{12}$ elastic constant decreases
with pressure\cite{Hailing} and becomes
negative in the high pressure phase, indicative of an intermediate
valence system.\cite{wachter} 
The pressure induced semiconductor to metal transformation
has also been traced by
optical reflectivity studies,\cite{Kaldis} 
M\"ossbauer measurements,\cite{Coey} 
resistivity measurements,\cite{Sidorov} and 
inelastic X-ray scattering.\cite{Raymond}
Valence transitions can also be
brought about by alloying with other trivalent ions, such as 
Y, La, Ce  or Gd, into the SmS lattice.\cite{wachter,Jayaraman2} 
%at room temperature for SmS under pressure
%shows a sharp increase in isomer shift around 6 Kbar 
%accompanying the transformation from the black to golden phase\cite{Coey}.  
%and resistivity measurements.\cite{Sidorov}
%Experimental studies on the resistivity of Sm chalcogenides under pressure 
%reveal that the intermediate valence state is stabilised after phase
%transition at pressures of 0.65 GPa in SmS, 3.4 GPa in SmSe and 5.2 GPa in 
%the case of SmTe\cite{Sidorov}. 
%Recently, high-pressure X-ray diffraction studies have been 
%carried out on SmS, SmSe and SmTe upto 55 GPa in which 
%a pressure induced valence change is observed around 1.24, 7 and 5 GPa for 
%SmS, SmSe and SmTe, respectively.\cite{LeBihan} 
Similar valence instabilities are observed in
SmSe and SmTe,\cite{Jayaraman,Campagna,Sidorov,LeBihan}
which also crystallize in  the NaCl structure.
For these compounds the volume changes continuously, but anomalously,  
with compression (at room temperature).\cite{Chatterjee,LeBihan}
From the photoemission studies it is concluded that 
SmSe and SmTe at amibient pressure, like SmS,
are also of predominantly  $f^6$  character,\cite{Campagna}    
while the monopnictides show clear signals of pure $f^5$ ions.\cite{Pollak,Campagna}

The present paper addresses two issues of the SmX compounds. Firstly,
density functional based total energy calculations are used to investigate the ground
state valency of Sm. To validate the approach, calculated values  of lattice constants and 
(for the Sm chalcogenides) valence transition pressures are compared to
experiment. Secondly, electron spectral functions are calculated by combining the band
structure Hamiltonian of the total energy calculation with an atomic description of
the multiplets of the localized $f$ electrons.

The theoretical description of Sm compounds is a challenge due to the Sm $f$-electrons.
Conventional band structure calculations of SmS,\cite{Farberovich,Strange,Lu,Schumann} 
as implemented within
density functional theory\cite{kohn} in the local density approximation (LDA),
describe the 
$f$ electrons as narrow bands. 
A spin-orbit splitting of about 0.6 eV between occupied $4f_{5/2}$ and 
unoccupied $4f_{7/2}$ bands is found, and the valency transition can be
traced by the crossing of the occupied $f$ bands with the lowest $d$ conduction band
upon compression.\cite{Strange}
However, the bonding of the $f$ electrons when described as band states
is greatly overestimated, as the
calculated lattice constant of SmS from the LDA calculations is 7.6\%
smaller than the experimental value.\cite{Lu} 
To treat the $f$ electrons as localized,   Schumann et al.\cite{Schumann} 
applied
self-interaction corrections (SIC) to the six 
$4f_{5/2}$ states. This led to a
semiconducting ground state of  SmS, however with a too wide gap of charge transfer type,
from S $p$ to Sm $d$ bands, in contrast to the Sm $f$ to $d$ character inferred from 
experiment.\cite{wachter} 
%move?
The reason is that the SIC scheme does not 
position the energy of the localized $f$ states properly with respect to the band states. 
While the
formalism does lead to  some SIC $f$ eigen-energies (which formally are just Lagrange
multiplier parameters of the theory with no physical interpretation), 
these are poor representations of the 
physical removal energies due to the localized nature of the excitations.
To remedy this, subsequent work  of
 Lehner et al.\cite{Lehner} calculated the
spectral density of SmS using a multiband periodic Anderson model, in which  the $s$, $p$, and
$d$ states were treated as band states within LDA, while
the $4f$ states were treated as localized states including atomic multiplets. 
The calculated spectral density
was found to be in good agreement with the measured photoemission and inverse 
photoemission spectra. Recently, the electronic structure and optical spectra 
of SmS, SmSe and SmTe were calculated with the LSDA+U approach.\cite{Antonov}  
This scheme includes an effective Coulomb parameter to separate bands of occupied and unoccupied
correlated electrons. Some ambiguity exists as to how this is most appropriately 
done,\cite{Petukhov} but the implementation of Ref. \onlinecite{Antonov}
leads to good agreement with experimental optical spectra, although
multiplet effects still have to be considered separately.

In the present work the research efforts outlined above will be pursued further.
The trends of the electronic structures of all of the samarium pnictides 
and chalcogenides will be investigated.
We separate the issues of total energy and ground state determination from that of 
excited states properties.
The total energy is calculated  using the self-interaction corrected  local-spin density (LSD) 
approximation.\cite{zp} When applied to rare earth systems this scheme
may describe the electron states as
either localized or delocalized.\cite{nature} For Sm compounds the relevant scenarios would have
either 5 or 6 localized $f$-electrons on each Sm atom, 
corresponding to a trivalent or divalent Sm ion, respectively,
and by comparison of the total energy the ground state configuration can be determined.
The  details of the SIC-LSD
approach and its implementation will be presented in Section II.A of this paper. 
The SIC-LSD method is %a total energy method, i. e.,
%  it computes within its approximations the total energy of a given
% system with a given configuration of rare-earth ions. 
% Thus, the method is 
primarily suitable for calculating 
ground state configurations, crystal structure and  equilibrium  lattice constants. 
Results for these parameters are presented in
Section III, including the discussion of pressure induced valency transitions. 
The photoemission spectra of SmX compounds are discussed by comparison to
calculated spectral functions. 
The density of states from a total-energy calculation cannot be directly compared with
experimental photoemission spectra. This is particularly clear in systems with 
partly filled localized
shells, where photoemission spectra reflect the multiplet structure
of the final state. To include these effects, in the present work
the LDA band Hamiltonian is combined with
additional atomic multiplet information, which is calculated in an isolated  
reference atom in the Hubbard-I approximation.\cite{Hubbard}
%, similar to the approach of Ref. \onlinecite{Lehner}. 
 The technical details of the present
implementation\cite{LDA++} are discussed in Section II.B.
In Section III the calculated spectral functions are presented and 
discussed in relation to experimental photoemission spectra.
Finally, Section IV  contains the conclusions
of the present work.

\section{Theory}

Two important theoretical tools are employed in this work for the investigation of the 
electronic structure of SmX compounds. The ground state total energies are calculated with the
self-interaction corrected local-spin-density method,\cite{zp,brisbane}
 while the spectral functions are
evaluated with the 'LDA++Hubbard I' atomic corrections method.\cite{LDA++}

\subsection{The SIC-LSD total-energy method}

The starting point is the total energy functional of the
LSD approximation, which is renowned for its chemical accuracy
in describing conventional weakly correlated solids.\cite{gunnarsson} 
The essential approximation is the parametrization of
the contribution to the total energy, $E_{xc}$, due to exchange and correlation 
effects among the electrons, by a simple term based  on the homogeneous
electron gas energetics.
To facilitate an accurate description of the localized $f$ electrons of
 rare earths, the self-interaction correction is introduced.
This correction constitutes a negative energy
contribution for an $f$-electron to localize, 
which then competes with the band formation energy gained by the
$f$-electron if allowed to delocalize and hybridize with the available conduction states. 
The SIC-LSD reduces the overbinding of the LSD approximation for narrow band states.
Specifically, the SIC-LSD\cite{zp} total energy functional is obtained from the LSD
as:
\begin{equation}
E^{SIC}=E^{LSD}-\sum_{\alpha }^{occ.}\delta _{\alpha }^{SIC}+E_{so},
\end{equation}
where $\alpha $ labels the occupied states and $\delta _{\alpha }^{SIC}$ is
the self-interaction correction for state $\alpha $. As usual, $E^{LSD}$ can
be decomposed into a kinetic energy, $T$, a Hartree energy, $U$, the
interaction energy with the atomic ions, $V_{ext}$, and the exchange and
correlation energy, $E_{xc}$.\cite{kohn} 
The self-interaction is defined as
the sum of the Hartree interaction and the exchange-correlation energy for
the charge density of state $\alpha $: 
\begin{equation}
\delta _{\alpha }^{SIC}=U[n_{\alpha }]+E_{xc}[n_{\alpha }].
\end{equation}
For itinerant states, $\delta _{\alpha }^{SIC}$ vanishes identically. 
For localized (atomic-like) states the self-interaction may be appreciable,
and for the free atoms the SIC-LSD approximation was demonstrated to be more accurate
than LSD.\cite{zp} In Sm compounds, the self-interaction correction term
is of the order $\delta_{\alpha}\sim 80$ mRy per $f$-electron.
Furthermore, the spin-orbit term is added:
\begin{equation}
E_{so}= \langle \xi(\vec{r})\vec{l}\cdot\vec{s} \rangle .
\end{equation}
We employ the atomic spheres approximation, whereby the crystal volume is divided into 
slightly overlapping atom-centered spheres of a total volume equal to the actual volume.
In (3), the angular momentum operator, $\vec{l}=\vec{r}\times\vec{p}$, 
is defined inside each atomic sphere, with $\vec{r}$ given as the position vector
from the sphere center.
Other relativistic effects are automatically included by solving the scalar-relativistic
radial equation inside spheres.

The $E^{LSD}$ functional includes the spin-polarization energy, which is a very important
contribution in the energetics of SmX compounds. The Sm chalcogenides are
non-magnetic,\cite{wachter} while the pnictides are antiferromagnetic with very low N\'eel
temperatures.\cite{beeken1,beeken2} 
In the present work the spin-polarized energy functional is not used to describe
interatomic magnetism but rather to describe the intra-atomic exchange interactions. 
By Hund's
first rule, the total spin of the $f^5$ or $f^6$ ion will be 5/2 or 3, and the ensuing
energy gain is reasonably well described with the LSD exchange energy. In this spirit,
the delocalization-localization transitions in elemental Ce\cite{glotzel} and Am\cite{skriver}
were described in the LSD approximation as the points of onset of spin polarization
of $f$ bands. In the SmX compounds the
effects of a spin-polarized $f$-shell have negligible influence on the occupied non-$f$
band states,
which are predominantly ligand $p$ like. An important issue, discussed recently for Eu
monochalcogenides,\cite{euch} is whether ferro-magnetically ordered Eu $f^7$ ions can
induce a spin splitting of the conduction band states, which can be exploited in
spin-filtering. 

Of   Hund's second and third rules, the latter is governed by the spin-orbit
interaction, Eq. (2), which also induces the formation of an orbital moment of the
$f$-shell. The contribution to the total energy due to the second
Hund's rule is, however, not fully   accounted for by the
spin-orbit interaction alone. The dominating contribution originates from the 
energy gain by forming the eigenstates of the two-electron interaction $1/r_{ij}$,
the tetrad effect.\cite{jorgensen} Little is known about how to implement these effects
in self-consistent solid state calculations. The orbital polarization scheme\cite{brooks1}
 has had some success in
describing the residual orbital moments of itinerant magnets like
UFe$_2$.\cite{brooks}

The advantage of the SIC-LSD energy functional 
in Eq. (1) is that different valency scenarios can be
explored by assuming atomic configurations with different total numbers of
localized states. In particular, these different scenarios constitute local
minima of the same functional, $E^{SIC}$ in Eq. (1), 
and hence their total energies may be compared. The
state with the lowest energy defines the ground state configuration. Note,
that if no localized states are assumed, $E^{SIC}$ coincides with the
conventional LSD functional, i.e., the Kohn-Sham minimum of the $E^{LSD}$
functional is also a local minimum of $E^{SIC}$. The interesting question
is, whether competing minima with a finite number of localized states exist.
This is usually the case in $f$-electron systems\cite{nature}
 and some 3d transition
metal compounds,\cite{tmo} where the respective $f$ and $d$
orbitals are sufficiently confined in space to benefit appreciably from the
SIC.
The SIC-LSD energy functional in Eq. (1) is a true density functional, as discussed in Ref.
\onlinecite{comment}.

The SIC-LSD still considers the electronic structure of the solid to be
built from individual electron states, but offers an alternative description
of the single-electron states to the Bloch picture, namely in terms of
periodic arrays of localized atom-centered states ({\it i.e.}, the
Heitler-London picture in terms of Wannier orbitals). Nevertheless, there
still exist states which will never benefit from the SIC. These states
retain their itinerant character of the Bloch form, and move in the
effective LSD potential. The resulting many-electron wavefunction will 
consist of both localized and itinerant states. In
contrast to the LSD Kohn-Sham equations, the SIC electron states, minimizing 
$E^{SIC}$, experience different effective potentials. This implies
that to minimize $E^{SIC}$, it is necessary to explicitly ensure the
orthonormality of the one-electron wavefunctions by introducing a Lagrangian
multipliers matrix. Furthermore, the total energy is not anymore invariant
with respect to a unitary transformation of the one-electron wavefunctions.
Both of these aspects make the energy minimization more demanding to
accomplish than in the LSD case. The electron wavefunctions are expanded in 
linear-muffin-tin-orbital (LMTO) basis functions,\cite{OKA}
 and the energy minimization problem becomes a
non-linear optimization problem in the expansion coefficients.
Further details of the present implementation
can be found in Ref. \onlinecite{brisbane}.

\subsection{Spectral Functions}

Photoemission spectroscopy is a powerful technique for investigations of rare earth 
compounds.\cite{Campagna2}
Several studies of the electronic structure of Sm chalcogenides and pnictides by
this technique have been reported.\cite{Campagna,Pollak,Oh,Chainani} The 
photoemission technique probes the
elementary electronic excitations of the material, revealing energy-resolved   and -
through cross section variations - angular momentum
decomposed information. The physical excitation
spectrum deviates substantially from the density of states (DOS) of the one-particle states 
exploited for the minimization of the SIC-LSD total energy.
Strictly speaking, the DOS contains information on the   one-particle band 
states that build up the SIC-LSD
ground state of the SmX compounds. In view of Janak's theorem,\cite{janak} the
LSD eigen-energies are good approximations to total energy differences, 
provided the quasiparticle in question is of extended character. 
(An unknown shift exists, though, between addition and
removal energies.\cite{PLSS} )
Therefore, for excitations which are {\it extended}, like the ideal Bloch waves
of band theory, the LSD eigen-energies compare reasonably
well with the physical excitation energies. However, 
often   the photon impact   creates {\it localized} excitations.
In the SIC-LSD approach one cannot expect the eigen-energies of the
localized $f$-electrons to have a straightforward physical interpretation.
The true many-body character of the electrons in the rare earth compounds
is borne out in the multiplet
structure of the photoemission spectra, which cannot be reproduced in an
independent-particle picture. This shortcoming is also encountered with 
the LDA and LDA+U approaches.
The multiplet structure arises from the two-electron interaction
$U_{ee}=\sum_{ij}'1/r_{ij}$, where the prime implies a sum over pairs $(i,j)$ of
electrons. This interaction splits 
the ionic $f^n$ configurations into terms characterized by good quantum numbers of
orbital, spin and total angular momentum, $L$, $S$ and $J$. In the ground state of the
solid usually only the lowest (Hund's rule coupled) term will be populated, maybe others
as well if close in energy (within $k_BT$, where $k_B$ is Boltzmann's constant,
 or a typical $f$ hybridization energy). But in the
final state of the photoemission process, several of the  $f^{n-1}$  terms may be
populated and  contribute significantly to the spectral features.

Recently, Lichtenstein and Katsnelson\cite{LDA++} devised a simple procedure to augment
LDA calculations to describe the multiplet features of partly filled shells. The main 
ingredient is an atomic selfenergy, $\Sigma^{atom}(\omega)$, extracted from the free ion
Green's function:
\begin{equation}
G^{atom}_{\mu\nu}(\omega)=\sum_{m,n} g_{mn}\frac{
\langle m|c_\mu|n \rangle 
\langle n|c^{\dagger}_\nu|m \rangle } {\omega+E_m-E_n+i\delta}.
\end{equation}
Here, $m$ and $n$ enumerate the ion multiplet states with energies $E_m$ and $E_n$, 
$c_\mu$ and $c^{\dagger}_\nu$
are, respectively,
 the annihilation and creation operators of single $f$-electrons, and $g_{mn}$ is a 
weight factor specifying the relevance of $m\leftrightarrow n$ transitions. In a thermal
environment,
\begin{equation}
g_{mn}=\frac{1}{Z} (e^{-\beta E_m}+ e^{-\beta E_n}),
\end{equation}
where $\beta = 1/k_BT$ and
$Z = \sum_m e^{-\beta E_m}$ is the partition function. 
The atomic levels $|m\rangle$ are found
by diagonalization of $U_{ee}$ in the $f^n$ manifold:
\begin{equation}
      H^{atom}_{ij}=<f^n;i|U_{ee}|f^n;j> +n(\epsilon_0-\mu_F),
\end{equation}
where $\epsilon_0$ is the bare $f$-level and $\mu_F$ is the Fermi level.
The matrix elements of $H^{atom}$ are expressed in
terms of Slater integrals, $F^k$, $k=0,2,4,6$
 and Gaunt coefficients,\cite{LDA++} and the Hamiltonian matrix is  easily 
diagonalized numerically. The Slater integrals are calculated using the $f$ partial waves
of the self-consistent LDA calculation.

From the atomic Green's function in Eq. (4) the atomic self-energy is extracted:
\begin{equation}
G^{atom}(\omega)=\frac{1} {\omega-\Sigma^{atom}(\omega)},
\end{equation}
and subsequently combined with the LDA Hamiltonian to give the solid state Green's function:
\begin{equation}
G^{solid}_{\bf k}(\omega)=\frac{1} {\omega-H^{LDA}_{\bf k}-\Sigma^{atom}(\omega)}.
\label{Gsol}
\end{equation}
The spectral function is evaluated as the imaginary part of the Green's
function of the solid:       
\begin{equation}
A^{solid}(\omega)=-\frac{1}{\pi} \text{Im}\sum_{\bf k} G^{solid}_{\bf k}(\omega).
\end{equation}
With care, the spectral function in Eq. (9) may be compared with experimental photoemission spectra.
The latter are also strongly influenced by matrix elements as well as secondary electron
losses and shake-up effects, all of which are not considered in the present work.
When combining the atomic and solid state electronic structure in this fashion, it is important
to ensure that no interaction is counted twice. In particular, since the atomic multiplets 
include the exchange interaction, the band Hamiltonian in Eq. (8) should not be spin-polarized.
Also, the LDA Hamiltonian includes the mean Hartree potential, which should therefore also be
subtracted in Eq. (6), but this correction is indistinguishable from  the Fermi-level
adjustment necessary to embed the atom in the solid. This Fermi-level adjustment is the
only fitting parameter of the scheme. If more than two $f^n$ configurations are compared, in
practice also a rescaling of the first Slater integral, $F^0$, is necessary to achieve results 
comparable to experimental data. This is due to screening in the photoemission process, which
in general renders bare Coulomb interaction too large (by a factor $\geq 2$). A completely
analogous screening effect is considered in the LDA+U approach.\cite{Antonov}
Lichtenstein and Katsnelson
also invoked screening effects of the higher Slater integrals for the interpretation of
TmSe photoemission spectra,\cite{LDA++}
but this was not considered in the present work.
The present approach is similar in spirit to, 
but differs considerably in details from, the approach
of Lehner {\it et al.} employed for SmS.\cite{Lehner}

\section{Results and Discussion}

\subsection{Cohesive Properties}

The calculated total energies as functions of unit cell volume are shown in figures 1 and 2, 
for SmAs and SmS, respectively. In each figure two curves are drawn, corresponding to
 the two cases of
five or six localized $f$-electrons on each Sm ion.  In SmAs the lowest energy is found for
Sm($f^5$) ions, while for SmS the lowest energy is found when assuming localized 
Sm($f^6$) ions.
The energy differences between the 
$f^5$ and 
$f^6$ minima are $\Delta E(f^5-f^6)=$ -74 mRy for SmAs and +15 mRy for SmS. The positions 
of the minima and the curvatures of the 
$f^5$ curve for SmAs and  the
$f^6$ curve for SmS yield the theoretical equilibrium lattice constants and bulk modulii,
which agree well with experimental values, cf. Table I.

To investigate further the electronic structures of 
the ground states found for SmAs and SmS
we show in Fig. 3 the DOS 
for these two compounds. 
For both SmAs and SmS, one
observes that the ligand $s$- and $p$-bands are 
completely filled. In SmAs, two narrow Sm 
$f$-resonances are situated just above the Fermi level. 
These are the $f$ majority spin bands for the two
$f$ orbitals which have not been treated as localized. 
The minority $f$ bands appear further above the Fermi level.
In SmS, similarly, there is one $f$ majority 
orbital which is not localized and therefore
appears as a narrow spin-up $f$ band, in this case 
$\sim 3$ eV above the conduction band edge. As a consequence
the density of states for SmS reveals a gap of $1.3$ eV.  
This is similar to the gap found in Ref. \onlinecite{Schumann} ($\sim 2$ eV)
and in agreement with the fact that SmS by experimental evidence
is a semiconductor. However, the gap in the DOS curve in Fig. 3(b) does not
correspond to the physical gap of SmS, which is of 
$f\rightarrow d$ character and only $\sim 0.15 $ eV in magnitude.\cite{wachter} 
The DOS curves  of Fig. 3 do not show the position 
of the localized states which, as discussed in Section II.B, are not well
represented by their respective SIC-LSD eigen-energies. 
Instead, the localized states can be calculated by total energy differences between the
ground state and a state with one $f$ electron removed.%\cite{norman}
 To achieve this  a supercell consisting of four SmS formula units was considered,
and the total energy calculated  for either all Sm atoms in the $f^6$ configuration,
or three Sm atoms in the $f^6$ configuration and one Sm atom in the $f^5$ configuration.
The missing electron in the latter case is artificially compensated by a uniform 
positive background, in accord with the scheme often employed for charged
impurities in semiconductors.\cite{semi}
With this approach 
the lowest $f\rightarrow d$ transition is found to occur at 0.58 eV 
below the conduction band edge in the DOS of Fig. 3(b). This  is in considerably better
agreement with the experimental band gap,
albeit still too large, which could be an effect of the limited size of
the supercell.

In Fig. 3(c), the density of states of the high pressure ($f^5$) 
phase of SmS is shown. Compared to the situation in
Fig. 3(b) one more electron per formula unit is available for 
band formation.
At the same time, due to the reduced screening of the Sm nuclear charge,
 the unoccupied $f$ bands move down in energy. 
The conduction states are being filled up to the position of the $f$ resonance,
which pins the Fermi level.
The inset shows that the first of the two
spin-orbit split majority $f$ bands becomes partly occupied.  
This means that each Sm ion is in a
configuration of mixed 
$f^5$ and
$f^6$ character, which we interpret as the SIC-LSD representation 
of an intermediate valence state,
in accord with the conventional wisdom for the golden phase of SmS.\cite{wachter}
The representation of the intermediate valence state, as 
built from a hybridized band on top of an array of
localized 
$f^5$ ions, may be a too simplistic description of the 
true quantum mechanical ground state of intermediate
valence,
yet due to the variational property % in the spirit 
of density functional theory it still leads 
to a good estimate of the binding energy.
The lattice constant at the minimum of the SIC-LSD 
energy curve in Fig. 2, calculated for the high pressure phase, $a=5.69$ \AA,
is in good agreement with the lattice constant obtained for 
the SmS golden phase, $a=5.70$ \AA\cite{wachter} or $a=5.65$ \AA.\cite{LeBihan}
We note, however, 
that the calculated energy balance between the 
$f^5$ and
$f^6$ phases in Fig. 2  is at variance with the 
occurrence of the isostructural transition in SmS already at 
$\sim 0.65$ GPa.\cite{Benedict1} The  common tangent 
construction applied to figure 2 would lead to a transition pressure
of $\sim 6.5$ GPa, {\it i. e.} ten times  higher. 
It is quite conceivable, though,  that a more accurate description of the
intermediate valence state would lead 
to higher binding energy for this phase, and hence to a lower transition pressure.
With increasing compression in the high pressure $f^5$ phase the $f$ band gradually
depopulates, at a rate of $dn_f/d \ln V\sim 0.5$, in accord with the experimental
observation of increasing valency of Sm in the golden phase of SmS with 
pressure.\cite{Barla}

From plots similar to those in Figs. 1 and 2 we can deduce the 
$f^5$-$f^6$ energy difference for all SmX compounds.
Figure 4 illustrates the trends in this quantity.
The calculations reveal a strong preference of 
the $f^5$  configuration in the early pnictides, with the energy difference of 
135 mRy per formula unit in SmN. For the heavier pnictide ligands, the $f^6$ configuration
becomes more and more advantageous, and for Bi it is only 6 mRy higher than the trivalent
configuration. Moving to the chalcogenides, already in the Sm monoxide the $f^6$ 
configuration is found to be most favorable, by 6 mRy, and in SmS by 15 mRy.
Hence, the SIC-LSD total energy predicts a Sm valency transition between the pnictides
and the chalcogenides. 
This  is not in complete agreement with   the experimental picture,
according to which the divalent and intermediate-valent states
 are almost degenerate in SmS, while SmO
is trivalent and metallic.\cite{smo,Krill} Thus,  it appears that
the SIC-LSD total energy functional overestimates the tendency to form the
divalent configuration of Sm, by approximately 15 mRy, in SmS. 
Assuming this error is similar for all SmX compounds,
this would imply that the calculated 
energy balance curve in Fig. 4 lies too high by approximately 15
mRy. Therefore, the figure also includes an indication of how the energy difference
behaves when a $\sim 15$ mRy correction is applied (dashed line). This 
switches the balance in favor of trivalency for SmO, which seems to be
in better agreement with experiments,
both with respect to the
lattice constant and metallicity of SmO.\cite{smo,Krill}
In the systematic study of the rare-earth metals and sulphides in Ref. \onlinecite{nature}
a similar uniform calibration (of 43 mRy)
was applied to the trivalent-divalent energy difference
of all the rare earths and rare earth sulphides,
in order to account for the experimentally observed valencies. The different size of the
calibrating energy
shift can to  a large extent be traced to the neglect of spin-orbit coupling in
Ref. \onlinecite{nature}, and the inclusion of this effect in the present work. In conclusion, 
the SIC-LSD total energy functional predicts correctly the trends in trivalent-divalent energy
difference through the samarium pnictides and chalcogenides, but fails on a
quantitative scale of the order of 15 mRy. Such an error is quite reasonable given that 
the functional does not contain any explicit
contribution from the formation of atomic multiplets
(the tetrad effect\cite{jorgensen}), which would lead to larger energy-lowering
 for an $f^5$ ion than for an $f^6$ ion.

The total energies as a function of volume in the Sm monopnictides and chalcogenides
are used to study the basic ground state properties such as equilibirium lattice constant
and bulk modulus. The calculated properties are in excellent agreement with the experimental 
values which are given in Table I. All lattice constants agree within $\sim 1 \% $ with 
experiment.
Figure 5 compares the calculated lattice constants of the SmX 
compounds with experimental values and the equilibrium lattice constants of the competing
valency configurations.

In view of the intermediate valence interpretation of the high pressure $f^5$ scenario of SmS
depicted in Fig. 3(c), one may wonder if the occurrence of the sharp resonance just above the
Fermi level in SmAs in Fig. 3(a) does not also imply a 
slight admixture of $f^6$ into the ground
state of this compound. Integrating the resonance band, one indeed finds
$\sim 0.11$ occupancy of this band, {\it i.e.} SmAs is found     to have 
a slightly mixed valent 
character. The admixture        of $f^6$ character increases to 0.17 in SmSb and 0.30 in SmBi.
Neither the experimental photoemission
spectra of SmAs nor SmSb show distinct features of $f^6$ 
character,\cite{Pollak,Campagna}
but on the other hand
it is unclear whether a $\sim 10$ \% admixture of $f^6$ character could be 
firmly excluded from the spectra. We are not aware of experiments on SmBi, but the present
prediction is that a significant fraction of $f^6$ should be present in this compound.
The proximity of the $f^6$ states is    also probed by doping experiments starting
from the pnictide and alloying with chalcogenides.\cite{Von,beeken1,beeken2} 
For SmAs, doping with
S or Se does not indicate valence instability for S concentrations up to 40 \%\cite{Von} or
Se concentrations up to 30 \%,\cite{beeken1} 
which suggests that in fact the extra electron of the chalcogen
 is not transferred into Sm $f$ states,
speaking against the presence of the unoccupied $f$ band right at the Fermi level.
On the other hand, when S is doped into SmSb the valency of Sm is seen to decrease already at
low concentrations of S.\cite{beeken2}
It seems that more    experimental as well as theoretical research,
including alloy calculations, is needed to elucidate this
issue further.

\subsection{Valence transitions}

As seen in figure 2 the divalent state, with
six localized $f$ electrons on each Sm ion, is the ground state of SmS, 
however, with compression the intermediate valent
phase with five localized $f$ electrons  and some additional
band $f$ electrons becomes more favorable. From a common tangent construction,
a transition  to the $f^5$ phase occurs, at a pressure of 0.1 GPa.  
This is accompanied by   a volume reduction of 11.1\%.
Note, that this transition pressure corresponds to the total energy curves calibrated
by the 15 mRy correction discussed
in Section III.A. %Without this correction, the calculated transition pressure would have been
%6.5 GPa, which is 10 times larger than the experimental value.\cite{Benedict}

In the case of SmSe a similar transition is calculated to occur at a pressure of
 3.3 GPa with a volume reduction of 9.8\%. 
Experimental evidence shows that  SmSe undergoes a continuous transition in the pressure range
of 2.6-4 GPa,\cite{tsiok} or 3-8 GPa.\cite{LeBihan} 
The present theory can only describe a discontinuous transition.
It has not been resolved whether the continuous volume change
 is due to the experiments being conducted 
at room temperature or is an intrinsic property of the quantum state of SmSe.
The calculated 
transition pressure and volume collapse are in reasonable agreement with the 
experimental data, cf. Table II,
the experimental volume jump being estimated by extrapolation of the divalent
$pV$-relation over  the anomalous region.
Note that the two recent experiments quoted above disagree considerably with respect to
the pressure range over which the transition occurs.

Similarly to SmSe, experimentally SmTe also exhibits a continuous divalent 
to intermediate valent transition with pressure. The present theory finds a discontinuous 
transition occurring at     6.2 GPa with a volume collapse 
of 8.4\%. Again, these values are in good quantitative agreement with the
experimental data, cf. Table II.

\subsection{Photoemission}

The spectral functions of SmX compounds are calculated as outlined in Section II.B. 
Figures 6 and 7 show the spectral functions for SmAs and SmS, respectively. The chemical
potential of the reference ion is chosen such that the ground state is 
$f^5({}^6H)$ for SmAs and $f^6({}^7F)$ for SmS
with an energy separation to the lowest $n-1$
excited levels ($f^4(^5I)$ of 4.0 eV for SmAs and 
$f^5({}^6H)$ of 0.8 eV for SmS, respectively), to coincide with
the experimental values for these energies.
The Slater integrals are almost equal for the two compounds,
$F^k=23.9, 10.6, 6.5$ and $4.6 $ eV, respectively,
 for $k=0,2,4,6$ (evaluated with the $f$-radial wave at
an energy given by the center of gravity of the occupied $f$-partial density of states).
However, for the direct Coulomb parameter, 
$F^0\equiv U$, a screened value of $F^0=7.1 $ eV was
adopted instead of the unscreened value quoted above.

The SmAs spectral function in Fig. 6 shows the four distinct peaks corresponding to the
$f^4(^5L)$, $L=D,G,F,I,$ final states in the photoemission process. 
These states agree well with the
three-peak structure observed by Ref. \onlinecite{Pollak}, at binding energies of
approximately -10.0 eV, -8.2 eV and -6.0 eV (presuming that
the $^5F$ emission is too weak to
lead to a resolvable peak). In the positive frequency region one observes the 
$^7F$ peak just above the Fermi level, in accord with the unoccupied majority
$f$-band in Fig.  3(a). 
The $f^6$ final states of $S=2$, which correspond to the unoccupied
spin down bands in Fig. 3(a), are situated further up in energy,
however now with a considerable spread due to the many
allowed multiplets. The position of the corresponding levels in the
reference atomic calculation are marked in the figure. 

The SmS spectral function is shown in Fig. 7. 
The spectrum is now characterized by the low binding energy three-peak structure,
which is also observed by several experiments,\cite{Campagna,Pollak,Chainani} at binding
energies -0.8 eV, -1.5 eV and -4.0 eV, and which is attributed to the $^6H$,
$^6F$ and $^6P$ final states.\cite{Campagna} 
The latter state     coincides with the S $p$-band, as also found in
the calculations. The results in Fig. 7 are similar to those obtained by
Lehner et al.\cite{Lehner}
Recent experiments\cite{Chainani} show traces of Sm $f^5$ emission in SmS
photoemission experiments, possibly also present in older works.\cite{Pollak} 
It is unclear, whether this
is due to small impurity concentrations or implies a more complicated ground state 
already for the
black phase of SmS. The present total energy calculations 
have found SmS to be a purely divalent system. 
It is well known that doping of SmS can lead to the intermediate valence phase,
characterized by photoemission spectra of both the high and low binding energy
type.\cite{Campagna,Pollak} By carefully tuning the chemical potential of the reference
atom in the present theory we can indeed obtain a mixed spectrum, corresponding to a
superposition of the spectral functions of Figs. 6 and 7, in good agreement with
the spectra recorded for SmAs-SmS alloys.\cite{Pollak} 
%{\bf Shall I include such a figure?}
The unoccupied states of SmS have been monitored with bremsstrahlung inverse spectroscopy.\cite{Oh}
The spectra reveal two broad structures, approximately 4.5 and 9 eV above the Fermi level,
which are in good agreement 
with the positions in Fig. 7 of the $^8S$ and $^6X$ features, respectively.

\section{Summary}    

The cohesive properties   of SmX compounds are well described by the local density
approximation to density functional theory provided  the self-interaction correction
is applied to obtain an improved description of the atomic-like $f$ electrons. The
bonding properties are quantitatively in agreement with experiment as evidenced by
accurate lattice constants for both the trivalent pnictides and the divalent
chalcogenides. Regarding  the energy
balance between the trivalent and divalent configurations of Sm in
the studied
solids ({\it i.e.} between localized $f^5$ and $f^6$ configurations), the 
SIC-LSD approach seems to underestimate the bonding in the localized $f^5$ configuration
by 10-15 mRy, which can be considered a minor error. However, for an accurate
description of the isostructural valence transitions induced by pressure it is a substantial
inaccuracy.
Correcting for this error we obtain good agreement with high pressure experimental 
results for SmS, SmSe and SmTe. The high pressure phase of the Sm chalcogenides is
described in the SIC-LSD one-electron picture as an array of Sm $f^5$ ions with an
additional partially occupied $f$-band, leading to a total $f$ occupation between 5 and
6. For SmO, this is found to be the ground state. A small  expansion of the SmO lattice,
corresponding to an effective negative pressure, would lead to a transition to the
divalent and semiconducting phase. This effect could be explored in SmO-SmS alloying
experiments.

The occurrence of multiplet effects in the photoemission experiments of SmX compounds 
is direct evidence that the simple one-electron picture does not suffice to
account for all physical characteristics
in these compounds. We demonstrated that the inclusion of local atomic 
correlation effects provides much improved 
 spectral functions, as seen      by the
close correspondence between the calculated
main peaks and experimental photoemission spectra. A certain degree  of fitting 
(of chemical potential and effective Coulomb interaction, $U$) 
goes into this procedure, which hence cannot
be considered as 'ab-initio' as the density functional based total energy calculations.
Therefore, substantial  further theoretical developments would be needed for a fully
 parameter free calculation of photoemission spectra. 

Concluding, this work has investigated the degree of intermediate valence in
Sm pnictides and chalcogenides as manifested in three physical properties,
namely the cohesive properties, the pressure characteristics and
the photoelectron spectroscopies.

\section{acknowledgements}

This work was partially funded by the EU Research Training Network
(contract:HPRN-CT-2002-00295) 'Ab-initio Computation of Electronic 
Properties of f-electron Materials'. Support from the Danish Center for Scientific
Computing is acknowledged. The authors V.K. and G.V. acknowledge the Max-Planck
Institute for financial support. 
The work of L.P. was supported in part by the Defense Advanced Research Project Agency 
and by the Division of Materials Science and Engineering, US Department of Energy, 
under Contract No. DE-AC05-00OR22725 with UT-Battelle LLC.

\newpage 

{\widetext

\begin{table}[t]
\label{a0}
\begin{ruledtabular}
\begin{tabular}{|c|cc|cc|}
Compound & \multicolumn{2}{c|}{Lattice constant (a.u)}   & \multicolumn{2}{c|}{Bulk modulus (GPa)} \\ 
         & Present & Expt.$^{a}$      & Present & Expt. \\ 
	 \hline
SmN  & 9.46   & 9.52   & 131   & - \\
SmP  & 10.99  & 10.88  & 68.7  & -  \\
SmAs & 11.18  & 11.16$^{f}$  & 67.6  & $84.2(3.5)^b$, 78.3$^i$ \\
SmSb & 11.90  & 11.84  & 46.7  & - \\
SmBi & 12.14  & 12.01  & 41.3  & - \\
SmO  & 9.41   & 9.34$^{g}$   & 106.6 & -  \\
SmS  & 11.25  & 11.25$^{f}$  & 53.4  & $42(3)^c$, $50.3^d$, $47.6(5.0)^e$ \\
SmSe & 11.70  & 11.66  & 53.4  & $40(5)^c$ \\
SmTe & 12.43  & 12.46  & 37.6  & $40(5)^{c}$ \\
SmPo & 12.64  & 12.71$^h$  &  $33.4$     &   -              
\end{tabular}
\end{ruledtabular}
\caption{Calculated lattice constant and bulk modulii of SmX compounds
in the NaCl  structure. The Sm ions are in the calculated ground state configuration of $f^5$
for the pnictides and SmO, and of $f^6$ for the other chalcogenides. \\
$^a$: Reference \onlinecite{pearson}, except where reference to other work is given. \\
$^b$: Reference \onlinecite{Shirotani}. \\
$^c$: Reference \onlinecite{Benedict1}. \\
$^d$: Reference \onlinecite{Hailing}. \\
$^e$: Reference \onlinecite{Kaldis}.  \\
$^f$: Reference \onlinecite{Pollak} at 110 K. \\
$^g$: Reference \onlinecite{Krill}.\\
$^h$: Sm$_{0.532}$Po$_{0.468}$.  \\
$^i$: Reference \onlinecite{Von}.
 } 
\end{table}

\begin{table}[b]
\label{Ptrans}
\begin{ruledtabular}
\begin{tabular} {|l|cc|cc|}
Compound & \multicolumn{2}{c|}{$P_t$(GPa)} & \multicolumn{2}{c|}{Volume collapse (\%)} \\
                                     
         & Theory & Expt.            & Theory & Expt. \\
                                     
\hline
SmS   & 0.1  & $0.65^a$,             $1.24^c$           & 11.1  & $13.5^a$, 13.8$^c$ \\
SmSe  & 3.3  & $\sim 4^a$, $3.4^b$, $3-9^c$,$2.6-4^d$   & 9.8   & $8^a$, $ 11^c$, $7^d$ \\
SmTe  & 6.2  & $2-8^a$, $5.2^b$, $6-8^c$,   $4.6-7.5^d$ & 8.4   &        $9^c$,$7^d$  \\
\end{tabular}
\end{ruledtabular}
\caption{Calculated isostructural  transition pressures, $P_t$ (in GPa),
and volume changes (in \%),
of Sm monochalcogenides. Experimentally, the transitions of SmSe and SmTe 
(at room temperature) are continuous,
while SmS exhibits a discontinuous volume change. 
The calculated transition pressures 
include the 15 mRy calibration
of the total energy calculated for the high pressure phase (see text).\\
$^a$: Reference \onlinecite{Benedict1}.  \\
$^b$: Insulator-metal transition of Reference \onlinecite{Sidorov}. \\
$^c$: Present author's estimates from figures of Reference \onlinecite{LeBihan} and 
$^d$: Reference \onlinecite{tsiok}.
The volume changes for SmSe and SmTe are obtained by extrapolation over 
the transition range.  }
\end{table}
}
\newpage

\begin{figure}
\begin{center}
\includegraphics[width=90mm,clip]{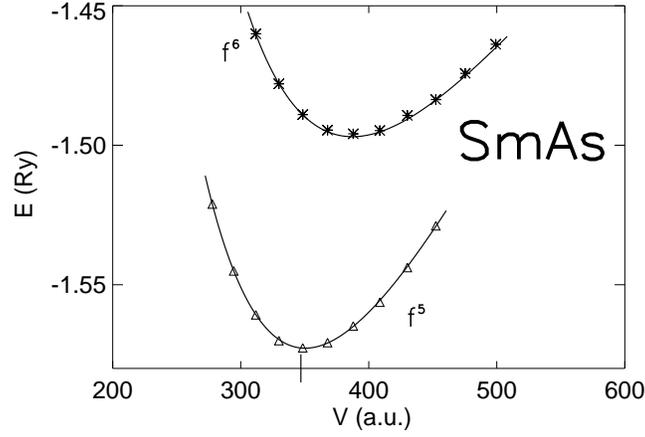}\\
\caption{ SIC-LSD total energy versus unit cell volume
 for SmAs. The two curves correspond to trivalent and divalent 
Sm ions, realized  by localized $f^5$ (triangles) and $f^6$ (stars) configurations, respectively.
The vertical bar on the $V$-axis marks the experimental equilibrium volume.}
\end{center}
\label{fig3self3}
\end{figure}

\begin{figure}
\begin{center}
\includegraphics[width=90mm,clip]{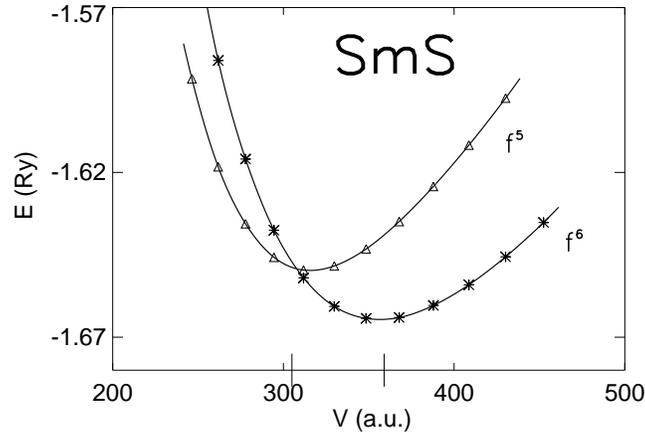}\\
\caption{ SIC-LSD total energy versus unit cell volume for SmS. 
The two curves correspond to trivalent and divalent 
Sm ions, realized  by localized $f^5$ (triangles) and $f^6$ (stars) configurations, respectively.
The data presented do not include the 15 mRy calibration for the $f^5 - f^6$ energy
difference (see text for discussion).
The vertical bars on the $V$-axis marks the experimental volumes of the black and golden
phases.}
\end{center}
\label{fig4self4}
\end{figure}
\newpage

\begin{figure}
\begin{center}
\vspace{-2cm}
\includegraphics[width=90mm,clip]{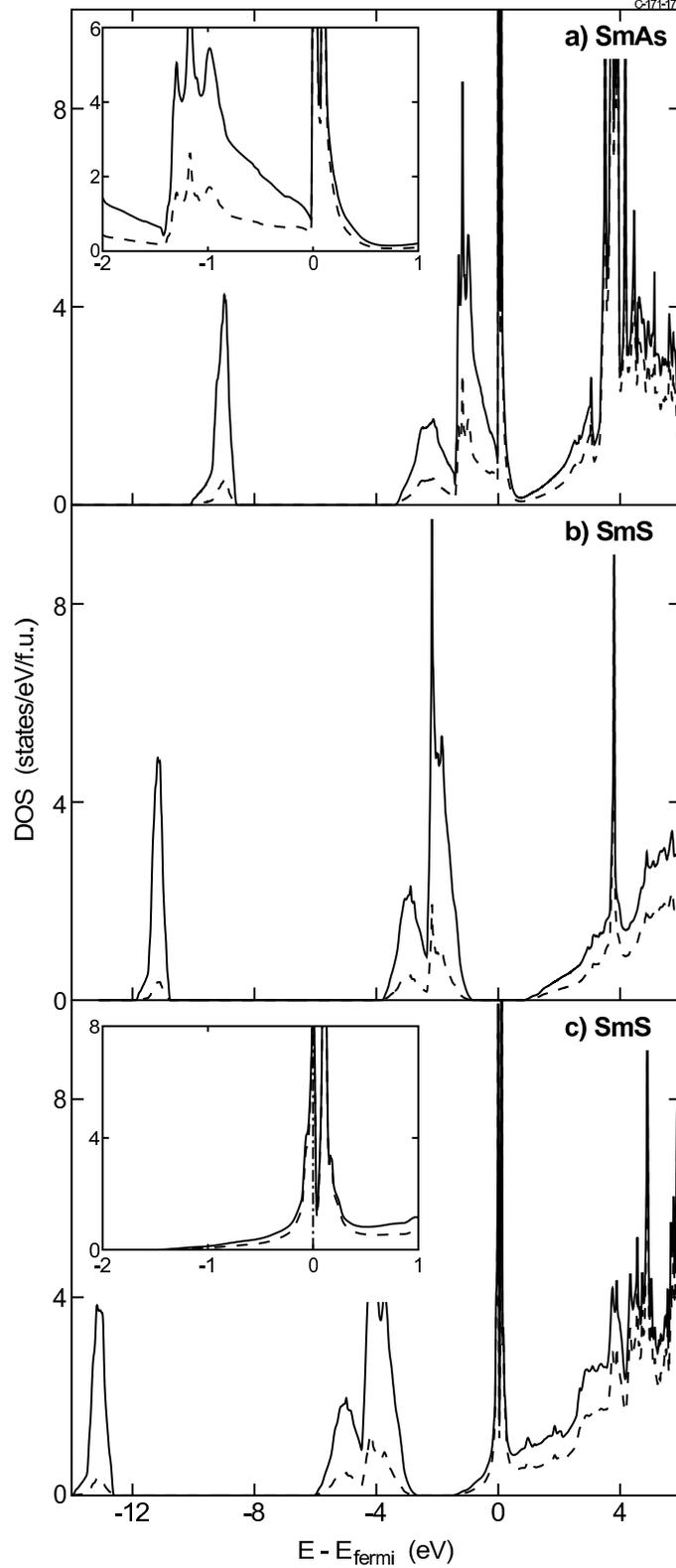}\\
\caption{ SIC-LSD densties of states (DOS) for: a) SmAs with localized $f^5$ Sm ions ($a=11.18$ a.u.),
b) SmS with localized $f^6$ Sm ions ($a=11.25$ a.u.), and 
c) high pressure phase of SmS with localized $f^5$ Sm ions ($a=10.75$ a.u.).
Energies are given relative to the Fermi level. 
The units of the DOS is states per eV and per formula unit.
The full line is the total DOS, while the dashed line shows the DOS projected onto the
Sm atom.
The insets in a) and c) show a blowup of the region in the vicinity of the Fermi level.
}
\end{center}
\label{dosfig}
\end{figure}

\begin{figure}
\begin{center}
\includegraphics[width=90mm,clip]{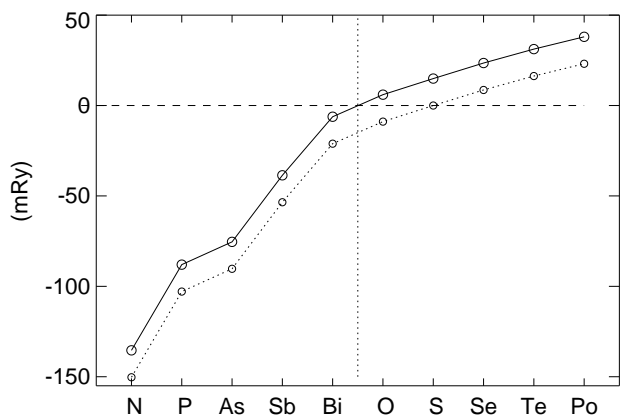}\\
\caption{ Trivalent-divalent energy difference (in mRy per formula unit) of Sm
compounds (full line).  
A negative sign implies that the trivalent state is favored. The dashed line marks the corrected
energy curve (see text).}
\end{center}
\label{fig1self1}
\end{figure}

\begin{figure}
\begin{center}
\includegraphics[width=90mm,clip]{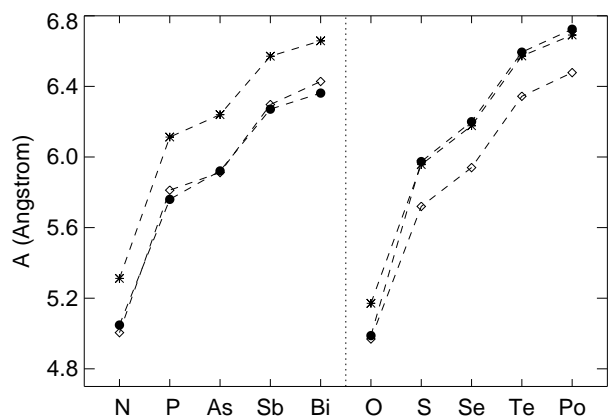}\\
\caption{ Comparison of experimental and theoretical lattice constants
(in \AA ngstr\"om)  of SmX compounds. Experimental values (see Table I) 
are marked with solid circles, while lattice constants 
calculated assuming a divalent (trivalent) Sm 
configuration are marked with stars (diamonds).}
\end{center}
\label{fig2self2}
\end{figure}

%\begin{figure}[t]
\begin{figure}
\begin{center}
\includegraphics[width=90mm,clip]{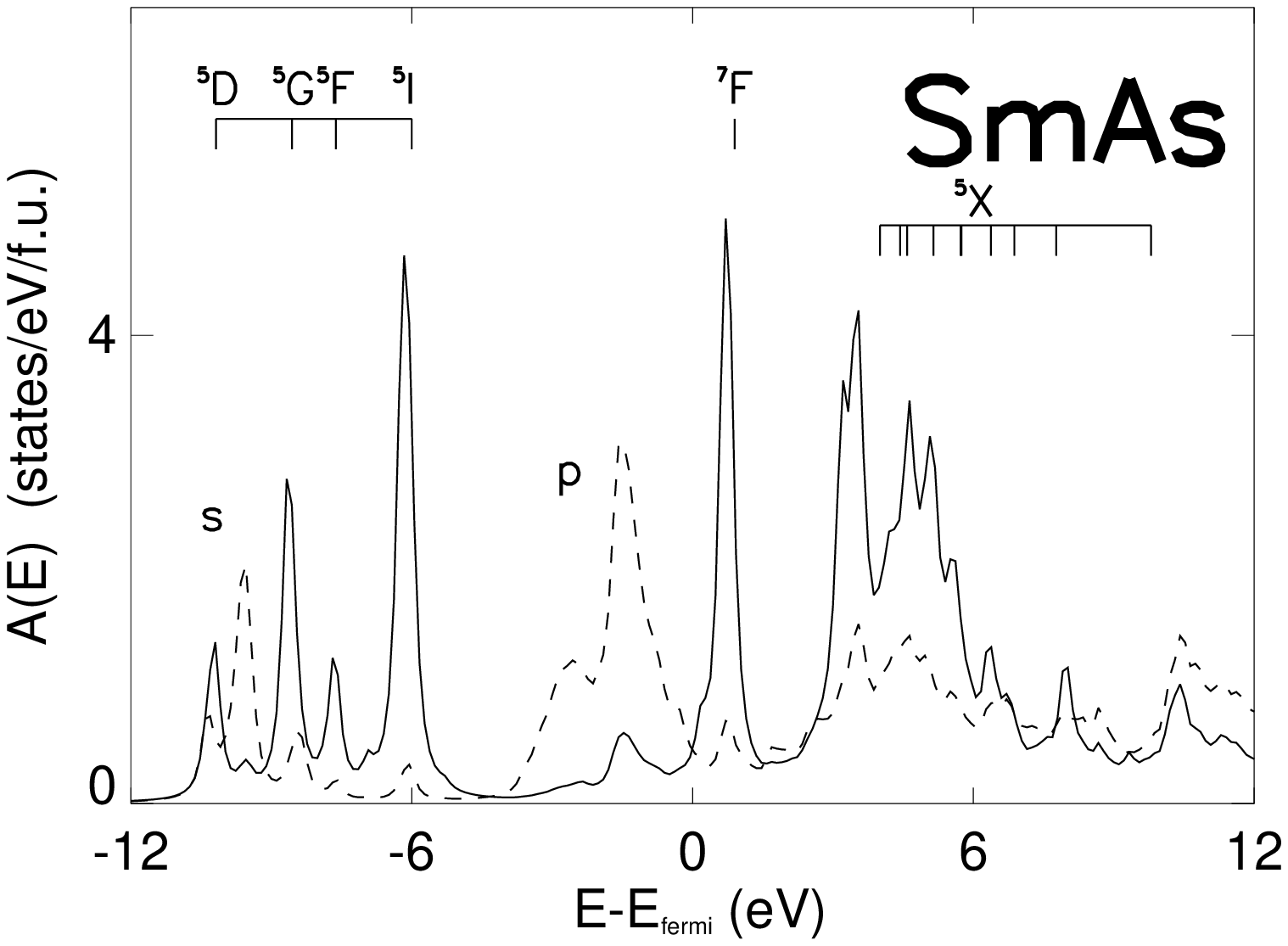}\\
\caption{ The calculated spectral function      of SmAs at equilibrium volume,
$a=5.91$ \AA. 
The full curve shows the $f$-contribution     and the dashed curve the non-$f$ 
contribution.  The energy is given relative to the Fermi level.
The main lines are characterized by their final state characteristics, either
As $s$, $p$-bands, or $f^{n\pm 1}$ multiplet term. 
}
\end{center}
\label{smas}
\end{figure}

\begin{figure}
\begin{center}
\includegraphics[width=90mm,clip]{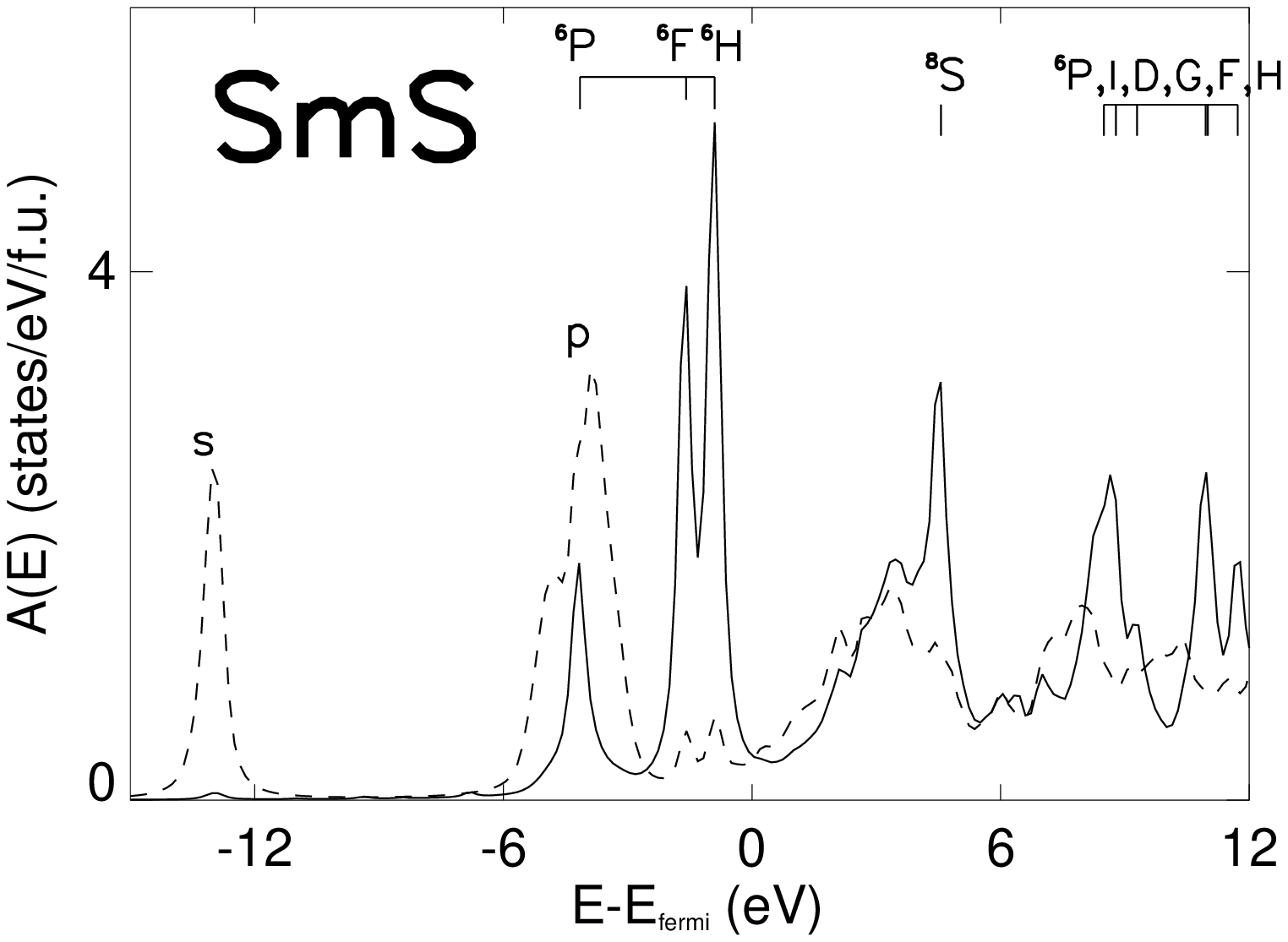}\\
\caption{ The calculated spectral function      of SmS (black phase, $a=5.95$ \AA). 
The full curve shows the $f$-contribution     and the dashed curve the non-$f$ 
contribution.  The energy is given relative to the Fermi level.
The main lines are characterized by their final state characteristics, either
S $s$, $p$-bands, or $f^{n\pm 1}$ multiplet term. 
}
\end{center}
\label{sms}
\end{figure}

%\begin{figure}[t]
%\caption{ Trivalent-divalent energy difference (in mRy per formula unit) of Sm
%compounds.  A negative sign means that the trivalent state is favored. The dashed line marks the corrected
%energy curve (see text).}
%\label{fig1}
%\end{figure}

%\begin{figure}[t]
%\caption{ Comparison of experimental and theoretical lattice constant of 
%SmX compounds. Experimental values [Pearson] 
%are marked with a solid ball, while lattice constants 
%calculated assuming a divalent (trivalent) Sm 
%configuration is marked with asterix's (diamonds).}
%\label{fig2}
%\end{figure}

\end{document}